\documentclass[amsfonts,amssymb,amsmath,prl,twocolumn]{revtex4}
\usepackage{times}
\usepackage{graphicx}
\usepackage{bm}
\usepackage{color}

\begin{document}

\title{Critical speed-up vs critical slow-down: a new kind of relaxation oscillation with application to stick-slip phenomena}

\author{Yves Pomeau$^1$ and Martine Le Berre$^2$ }

\affiliation{$^1$Department of Mathematics, University of Arizona, Tucson, USA.
\\ $^2$ Institut des Sciences Mol\'eculaires d'Orsay ISMO-CNRS, Univ. Paris-Sud, Bat. 210, 91405 Orsay, France.}

\begin{abstract}
The equations for the sliding of a single block driven by an elastic force  show numerically a fast and a slow step in their dynamics when a dimensionless parameter is very large, a limit pertinent for many applications.
  An asymptotic analysis of the solutions explains well the two sharply different steps of the stick-slip dynamics. The stick (slow) part takes place along a slow manifold in the phase space. But, in contrast with standard relaxation dynamics (of van der Pol type), the slow manifold is always formally attracting and the transition from slow to fast dynamics occurs because the slow dynamics has a finite time singularity breaking the assumption of slowness. This makes a new kind of relaxation oscillation. We show that the response of the stick-slip system to an external noise displays a progressive speed-up before the transition, in contrast with the well known critical slowing-down observed in the standard case.

\end{abstract}

\maketitle

\date{\today }

\section{Introduction}
\label{sec:Intro}

Many physical systems show relaxation oscillations (RO) with phases of slow motion interrupted by fast drifts.
Examples of RO in the real world are found in electro-mechanical devices\cite{gerard}, in purely electrical devices \cite{blondel}  and in self-oscillating circuits \cite{vdpol}. Besides these saddle-folding \cite{saddle} systems, it has been suggested by Brace and Byerlee in 1938 \cite{Brace} that earthquakes are RO of the stick-slip kind, appearing when two solids slide on each other. Here we compare the mechanism of RO in stick-slip models and in  saddle-folding ones for which a prototype
dealing with plane flow RO is the
 slightly extended van der Pol equation in the limit $\beta$ large, written as follows in its Li\'enard form
 \begin{equation}
 \left \{ \begin{array}{l}
\dot{y} =  x +a\\
\dot x = \beta^2 [x - \frac{x^3}{3}-y]\mathrm{.}
\end{array}
\right. \label{eq:1}
\end{equation}

 In this limit, the slow dynamics is a drift along the slow manifold (SM) of Cartesian equation $y = x - \frac{x^3}{3}$  interrupted at places (the folding spots) where this manifold is unstable against transverse perturbations.
 There are then two widely different time scales, the short one for the jumps away from the SM and the long one for the slow drift. In this class of models, the transition from fast  to slow motion occurs by dynamical saddle-node bifurcation \cite{canard}.
  Near the saddle-node transition (ie. at $x = \pm 1$ and $ y = \pm 2/3$) the loss of stability of the SM is generically the same as the one of a particle moving with friction in a potential changing slowly as a function of time, a  model recently investigated in view of predicting catastrophes \cite{SNL}.

This picture of RO
 in the strongly nonlinear limit with a slow manifold and quick jumps outside of this manifold seems to make the currently accepted "paradigm" for
  RO in
   dynamical systems. However, by studying the equations pertinent for the stick-slip dynamics we have found another scenario of RO, which behaves differently near the transition, with possibly important consequences on precursor phenomena.

Stick-slip phenomena occurring in solid friction are ubiquitous in real world, for example in geophysics and engineering. Earthquake rupture have been attributed to stick-slip instability observed in laboratory experiments  \cite {Brace}, \cite{ruina} and \cite{rice-tse}. It should be a good example of RO because the two phases (stick and slip ) take place with widely different speed.
Mathematical models for solid friction, as proposed by Dieterich \cite{dieterich} and by Ruina \cite{ruina}, make a fair representation of this physics, their numerical solutions showing  RO with two widely separated time scales. In the present paper we show that
 in this family of models the transition from slow to fast  does {\emph{not}} occur because the
 SM becomes repulsive, contrary to what happens in the strongly nonlinear regime of van der Pol-type models. Here the SM is attractive in the whole space. We show that the transition occurs because the trajectory on this manifold goes to infinity in finite time. Before the critical time $t_c$  the adiabatic approximation becomes invalid, and the real
 trajectory escapes from the SM.
 Therefore the stick-slip model of Dieterich-Ruina-Rice (DRR) makes  a new class of RO where the slow trajectory begins to accelerate a long time before $t_c$, whereas it slows down shortly before $t_c$ in the case of equation (\ref{eq:1}). It follows that the statistical properties of the responses to an external noise are completely different for the
  two classes of models. For stick-slip models
 the correlation time of the response decreases before $t_c$ (speed-up), whereas it increases (slowing-down) in the case of the standard saddle-node transition \cite{SNL}.

\section{DRR model equations}
\label{sec:DRReqs}
In the DRR models a "state" variable, denoted as $\theta$ below, describes the physical state of the surfaces sliding on each other (rugosities, asperities, etc.) and how it influences the friction. This state variable has its own dynamics, in agreement with the observations of time dependent solid/solid friction. Those models have been used to analyze the sliding of a block on a flat solid, the block of mass $M$ being driven by an external spring with the other end moving at constant speed $v_0$, this being possibly analogous to fault slip on the boundaries of Earth's crustal plate responsible of earthquakes \cite{ricetse}. We shall consider below the single-block problem.

We introduce first the DRR set of equations of motion of a single sliding block coupled with Dieterich-
Ruina rate-and-state dependent friction \cite{ruina}, \cite{ricetse} and discuss their solution in the limit where the inertia of the block is very small, the limit of the stick-slip sliding. This set of ODE's (ordinary differential equations) read

 \begin{equation}
 \dot{u} = v - v_0
 \mathrm{,}
 \label{eq:DR1}
 \end{equation}
 \begin{equation}
 \dot{v} = - \frac{1}{M}\left( k u + \theta + A\ln(v/ v_1)\right)
 \mathrm{,}
 \label{eq:DR2}
 \end{equation}
 and
  \begin{equation}
 \dot{\theta} = - \frac{v}{D_c}\left(\theta + B \ln(v/ v_1)\right)
 \mathrm{,}
 \label{eq:DR3}
 \end{equation}
 The first equation relates $u$, the position of the block, to its speed $v$. The constant $v_0$ is the difference of speed between the point holding the spring and the surface the block is sliding on. The force due to the external spring is proportional to $u$, with a spring constant $k$, and $M$ is the mass of the block. Moreover $\theta$, a contribution to the friction force, depends on the state of the surfaces facing each other, a function of the history of sliding, as given by the solution of the equation (\ref{eq:DR3}). The intermediate equation (\ref{eq:DR2}) is a way of writing the dynamics of the block under the effect of the pulling force proportional to $u$, of the friction, plus the state dependent part proportional to $\theta$. The quantity $(\theta + A\ln(v/ v_1))$ on its right-hand side is the friction force plus a  constant, absorbed into a constant displacement added to $u$.
 Even after addition of a constant friction (independent on $v$), the DRR equations cannot be valid for all values of $(u, \theta, v)$. Because $A$ is positive, the friction force is dominated at low speeds $v$ by the logarithm, and so becomes very large positive, in the direction of $v$, a negative friction  which is clearly nonphysical. Because the added constant is independent on the parameters used to write the equations above, one cannot tell, without knowing what is its value in a particular application, if the non physical regime of negative friction is reached or not for this case. We assume that, in the regime we consider, negative friction never happens. Note that in case of very low speed, Lapusta et al.\cite{Lapusta} propose to change the above relation $v/v_1= \exp(\tau_{f})$ between the velocity $v$ and the friction $\tau_{f}$, by the relation $v/v_1= 2 \sinh(\tau_{f})$, that changes $\ln(v)$ into $\ln(v+\sqrt{v^2+1})$.

 The velocity $v_1$ and the coefficient $A$, $B$ and $D_c$ are phenomenological quantities derived in principle from experiments.
One can absorb the ratio $v_0 / v_1$ by adding another constant to $u$. Using the scaled quantities $v/v_0$, $\theta/A$,$u/D_c$, $v_0 t/D_c$, one transforms the above equations into the mathematically convenient form \cite{these}

  \begin{equation}
 \dot{u} = v - 1
 \mathrm{,}
 \label{eq:DR1.resc}
 \end{equation}
 \begin{equation}
 \dot{v} = - \gamma^2 \left(u + \frac{1}{\xi}(\theta + \ln(v))\right)
 \mathrm{,}
 \label{eq:DR2.resc}
 \end{equation}
 and
  \begin{equation}
 \dot{\theta} = - v \left(\theta + (1+\epsilon) \ln(v)\right)
 \mathrm{.}
 \label{eq:DR3.resc}
 \end{equation}

This version of the DRR equations keeps three dimensionless parameters, $\gamma$, $\epsilon$ and $\xi$, related to the quantities $M$, $v_0$ and to the other phenomenological
 parameters  $A$, $B$ and $D_c$ derived in principle from experiments \cite{ruina}, \cite{ricetse}.
The geophysical literature  gives $\epsilon$ and $\xi$ of order one, and $\gamma$ large.
To give an order of magnitude, $\gamma$ is typically of order $10$ for ice flows
in Antarctica \cite{ice}, where the sliding phase lasts about $20$ minutes once a day, while for major earthquakes $\gamma \sim10^8$ \cite{rice-tse} with sliding phases of typically $5$ seconds occurring once every $200$ years.

\begin{figure}[htbp]
\centerline{$\;\;$
(a)\includegraphics[height=1.4in]{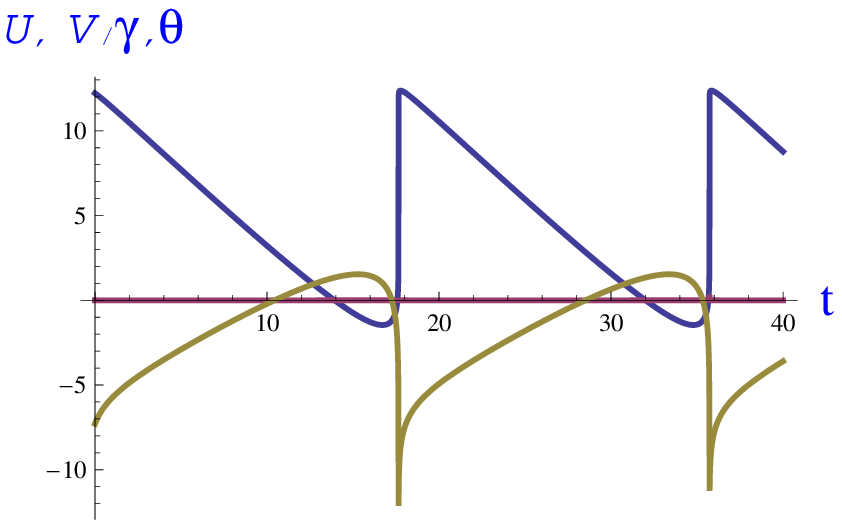}
 $\;\;$}
 \centerline{$\;\;$
(b)\includegraphics[height=0.75in]{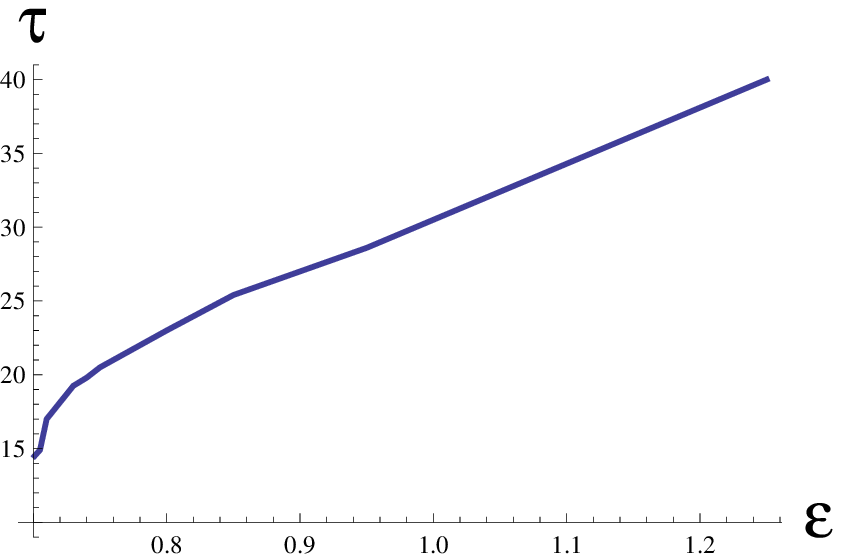}
(c)\includegraphics[height=0.75in]{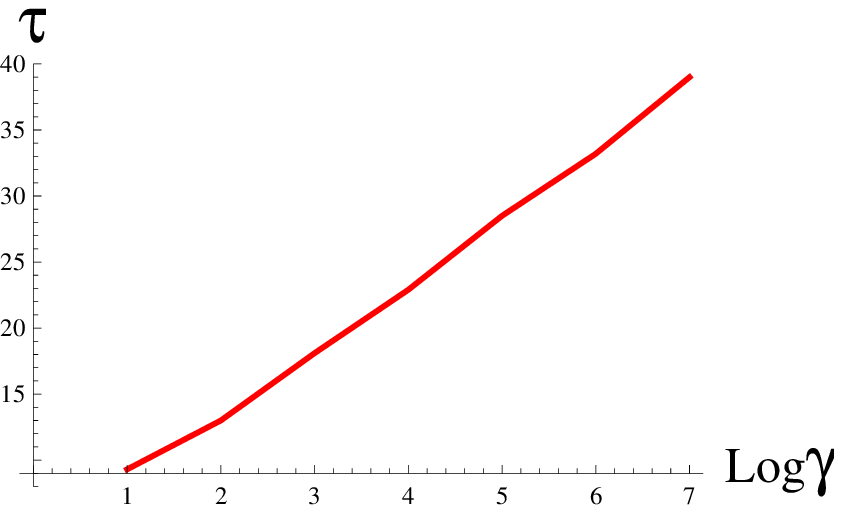}
 $\;\;$}
\caption{(a) Periodic solution of equations (\ref{eq:DR1.resc}-\ref{eq:DR3.resc}) for $\gamma=10^3$, $\xi=0.7$, $\epsilon=0.72$.
(b-c) period of the limit cycle for $\xi=0.7$, in (b) as function of $\epsilon$ for $\gamma=10^3$; in (c) as function of $\log_{10}(\gamma)$ for $\epsilon=0.72$.
}
\label{Fig:solution}
\end{figure}

 When the parameter $\gamma$ becomes very large and $\frac{\epsilon}{\xi}$ is above unity, the fix point ($u=\theta=0,v=1$) of equations (\ref{eq:DR1.resc})-(\ref{eq:DR3.resc}) undergoes a Hopf bifurcation, leading to a stable periodic solution,
 see figure (\ref{Fig:solution}-a). At threshold the period is
 \begin{equation}
  \tau_c =\frac{2\pi}{\sqrt{\xi}}
 \mathrm{.}
 \label{eq:tauc}
 \end{equation}
 The period $\tau$ increases noticeably with the control parameter, as illustrated by curves (b-c) in Figure (\ref{Fig:solution}) drawn for $\xi=0.7$, it reaches a linear dependence with respect to $\epsilon$ for given $\gamma$ , and increases nearly as $5\log(\gamma)$ for given $\epsilon$.

 In the case of earthquakes, the typical parameter values \cite{rice-tse} $v_0=30$ mm/year, $D_c/v_0 = 2.7$ years, $\xi=0.8, \epsilon=1, \gamma=10^8$  lead to  a  period  equal to $75$ years, or $\tau=28$ in units of equations (\ref{eq:DR1.resc})-(\ref{eq:DR3.resc}). More generally, considering earthquakes separated by $30$ to $200$ years, it gives the dimensionless period
 \begin{equation}
  \tau \sim 10 - 70
  \mathrm{,}
   \label{eq:fastime}
\end{equation}
which is the range of period values considered below in our numerical calculations.

 The striking point is
 the stick-slip character of the solutions with
  almost motionless long intervals (stick step) interrupted by short bursts of rapid displacement (slip step). In this limit the stiffness of the DRR equations makes them hard to solve numerically, so that most numerical simulations are done with $\gamma$ of order unity, unrealistic for earthquake modeling. Therefore is is important to carry an asymptotic analysis of the solutions of this set in the large $\gamma$ limit, what is done below.

  Moreover we check our
   theoretical results by a systematic numerical investigation of the solutions for  $\gamma$ in the range $10$-$10^4$, plus some calculations for larger values of $\gamma$ ( until $10^7$). This was performed on Mathematica, by
the "StiffnessSwitching" method using a pair of extrapolation methods as the default. The nonstiff solver uses the "ExplicitModifiedMidpoint" base method, the stiff solver uses the "LinearlyImplicitEuler" base method.

\section{Slow and fast regimes}
\label{sec:slow-fast}

 \subsection{stick step}

 Consider first the stick  regime, where the acceleration of the motion is very small,
 \begin{equation}
  \dot{v} \ll \gamma^2
 \mathrm{,}
 \label{eq:stick}
 \end{equation}
 although there is always some sliding as illustrated in figure (\ref{Fig:solution}-a).
  The time duration of this step
   is very close to the period $\tau$ of the limit cycle.

   The velocity $v$ and the acceleration $\dot{v}$ are smaller than one in the main part of the sticking stage, they both becomes  unity at time nearly equal to $t_{c}-1$.
   This defines the intermediate time scale $t_0$,
  the time interval separating the instant where $v(t)=1$ and the catastrophe time $t_c$ where the velocity is maximum. For the whole range of parameters that we investigated numerically, we have found the relation
   \begin{equation}
 t_0 \cong 1
 \mathrm{,}
 \label{eq:to}
 \end{equation}
 within the units of equations (\ref{eq:DR1.resc})-(\ref{eq:DR3.resc}).
  After reaching the value $v=1$,and $\dot{v}=1$ the velocity and the acceleration increase. At the end of the stick step, in the range defined by $ 1\ll v \ll\gamma$ and equation (\ref{eq:stick}), the solution is analytically tractable, see the section "matching slow and fast solutions" below.
   Note that in the case of earthquake, the intermediate time scale corresponds
      to few years ($2.7$ years in \cite{rice-tse}),
  that makes the DRR model completely different from the saddle-node model where the intermediate stage lasts a few hours only \cite{SNL}.

      During the whole sticking episode defined by equation (\ref{eq:stick}), the coefficient of $\gamma^2$ on the right-hand side of equation (\ref{eq:DR2.resc}) can be set to zero, that gives the relation
 \begin{equation}
\xi u + \theta + \ln(v) = 0
 \mathrm{,}
 \label{eq:DR2.resc.1}
 \end{equation}
 defining a surface  $\textit{S}$ of equation $ v(u,\theta) = e^{-(\xi u + \theta)}$ in the phase space, $(u,v,\theta)$. Putting the relation (\ref{eq:DR2.resc.1}) in
  equations (\ref{eq:DR1.resc}) and (\ref{eq:DR3.resc}), one get a set of two first order ODE's for $u$ and $\theta$ in which the large parameter $\gamma$ has disappeared,
   \begin{equation}
 \left \{ \begin{array}{l}
\dot{u} = e^{-(\xi u + \theta)}-1\\
\dot{\theta} = e^{-(\xi u + \theta)}[\epsilon \theta + (1+\epsilon)\xi u]
\mathrm{,}
\end{array}
\right. \label{eq:2D}
\end{equation}
 defining a \textit{2D flow} (two dimensional).

   From the point of view of the 3D dynamical system we started from, the 2D flow defined by equations (\ref{eq:2D}) takes place on the surface (manifold) $\textit{S}$,
    a reduction of the original 3D flow done by Gu et al. \cite{Rice-Gu} in the same limit.  The "stick" phase of the 3D dynamics fulfills equations (\ref{eq:stick}), and (\ref{eq:DR2.resc.1})-(\ref{eq:2D}).

     Let us show
    that in the 3D flow, the SM  is attracting. The surface $\textit{S}$  splits the 3D phase space $(u,v,\theta)$ into two subspaces, the one for which $\left(\xi u + \theta + \ln(v)\right)$ is positive and the one for which it is negative.
    In the large $\gamma$ limit,
    the dominant component of the time derivative of the 3D vector $(u, v, \theta)$ is $ \dot{v} $. The negative sign on the r.h.s. of equation (\ref{eq:DR2.resc}), ensures the that the SM is attractive from both subspaces, because taking $v$ very large positive one sees that $ \dot{v} $, the dominant component of the "velocity", defined by the set (\ref{eq:DR1.resc}), (\ref{eq:DR2.resc}) and (\ref{eq:DR3.resc}) is directed toward $v$ negative, and conversely for $\left(\xi u + \theta + \ln(v)\right)$ negative. This shows that the slow manifold is everywhere attracting in this limit $\gamma$ large.
    Therefore the classical "scenario" for van der Pol-type RO with an ejection out of a repelling SM cannot hold for DRR equations. We show later that in the DRR case, the 3D flow escapes from $\textit{S}$  because the 2D flow trajectory goes to infinity in \textit{finite time}, see Figures (\ref{Fig:solution2}).

\subsection{Slip step}
During the slip, both $v$ and $\dot{v}$ are "large", of order $\gamma$ and $\gamma^2$ respectively. Neglecting  $(-1)$ on the right-hand side of equation (\ref{eq:DR1.resc}) allows to eliminate the large parameter $\gamma$ from the equations by rescaling and addition. This is done by putting $v = \gamma V(T)$, $t = T/\gamma$, $u = U(T) + c_u \ln(\gamma)$ and $\theta = \Theta(T) + c_{\theta} \ln(\gamma)$. The quantities denoted as $T$, $V$, $\Theta$ and $U$ are of order 1 with respect to $\gamma$ as well as the constants $c_{\theta} \mathrm{,} c_u$. Setting to zero the part independent of $T$ in the equations derived from (\ref{eq:DR2.resc}) and (\ref{eq:DR3.resc}) one finds two algebraic equations for $c_{\theta}$ and $c_u$ with the solution
 \begin{equation}
 \left \{ \begin{array}{l}
c_{\theta} = -(1+\epsilon)\\
c_{u} =  \frac{\epsilon}{\xi} \mathrm{,}
\end{array}
\right. \label{eq:cut}
\end{equation}
 The functions $U(T)$,$V(T)$ and $\Theta(T)$ are solutions of a set of three ODE's free of the large parameter $\gamma$
  \begin{equation}
 \left \{ \begin{array}{l}
 U_{,T} = V\\
  V_{,T} = -  \left[ U+ \frac{1}{\xi}(\Theta + \ln(V))\right]\\
\Theta_{,T} = - V \left[\Theta +  (1+\epsilon) \ln(V)\right]\mathrm{,}
\end{array}
\right. \label{eq:UV}
\end{equation}
 where $X_{,T} $ is for $\frac{{\mathrm{d}}X}{{\mathrm{d}}T}$.

  \begin{figure}[htbp]
\centerline{$\;\;$
\includegraphics[height=1.2in]{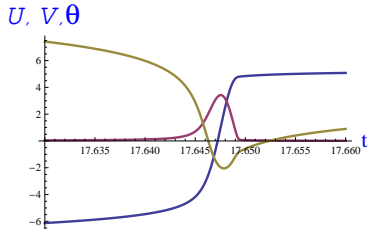}
 $\;\;$}
\caption{ Fast step. Part of the solution shown in Figure (\ref{Fig:solution}), close to the catastrophe, in units of equations (\ref{eq:UV})}
\label{Fig:solution3}
\end{figure}

This predicts that
 the slip lasts a (short) time $\delta t_{eqk} $ of order $1/\gamma$ although the sticking lasts a time independent on $\gamma$.  This scaling law is in excellent agreement with the numerical study of the full DRR equations for $\gamma$ large. We observed that, as a function of $t$, $v(t)$ peaks with a time half-width
  \begin{equation}
  \delta t_{eqk}\sim 2.5/\gamma
  \mathrm{,}
   \label{eq:fastime}
\end{equation}
   in a wide range of parameters, as illustrated in Figure (\ref{Fig:solution3}).

\section{matching slow and fast solutions}
\label{sec:matching}
\begin{figure}[htbp]
 \centerline{$\;\;$
 (a)\includegraphics[height=1.2in]{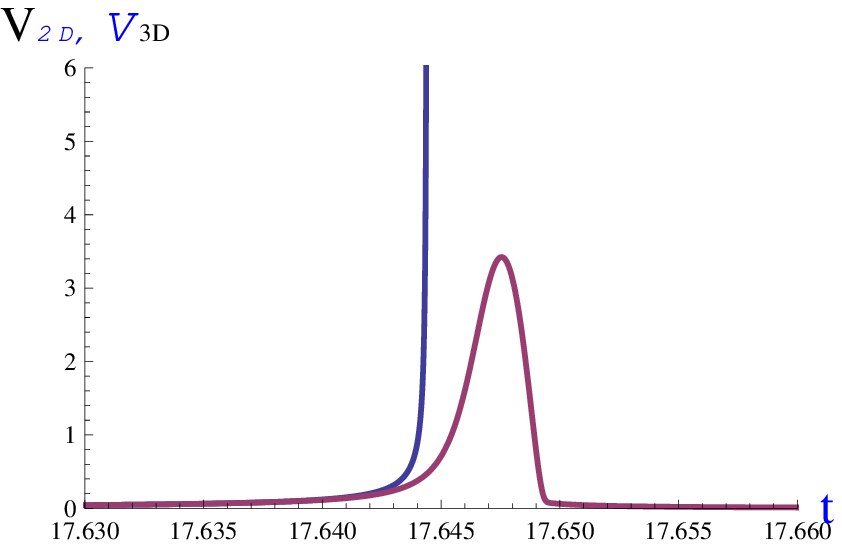}
  $\;\;$}
  \centerline{$\;\;$
 (b)\includegraphics[height=1.8in]{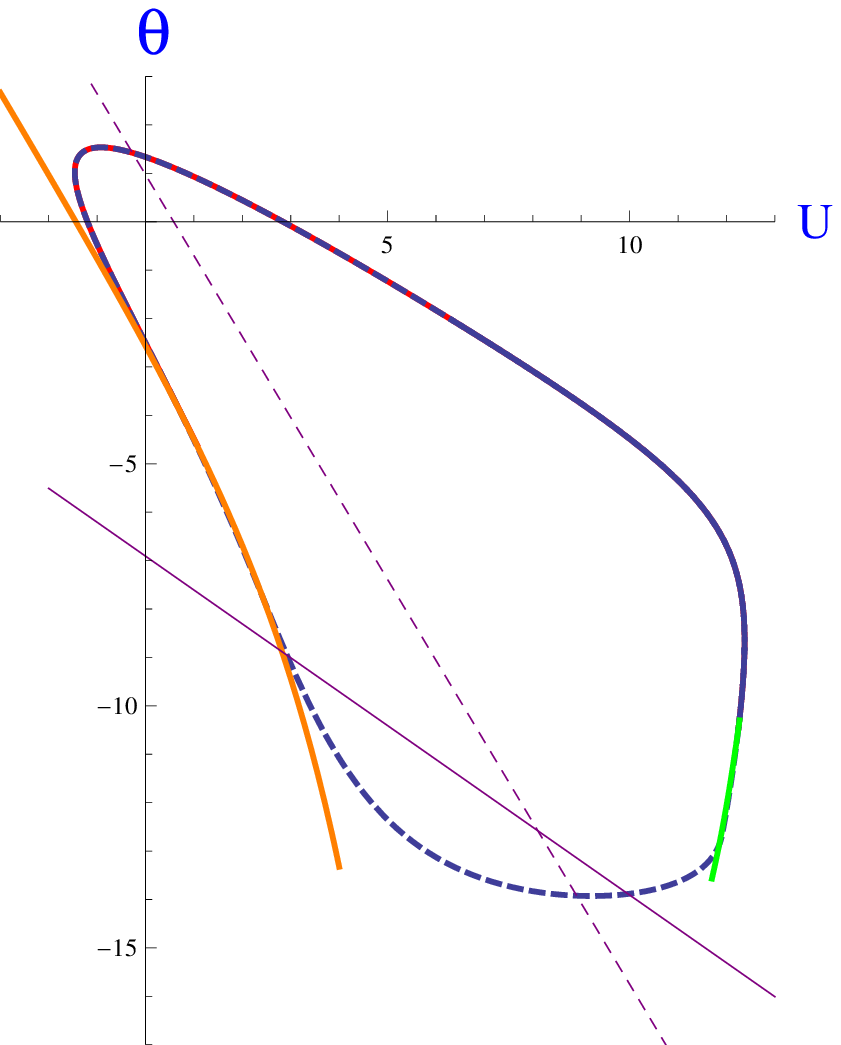}
  $\;\;$}
\caption{
(a)  $v(t)/\gamma$ for the 2D and 3D flows (blue and purple curves respectively) (b) Phase space ($u,\theta$). The 3D limit cycle (closed dashed blue curve) is superimposed with the 2D flow (solid red portion) on the SM. The orange and blue portions correspond to the two branches of matching solutions, equation (\ref{eq:raccord}).
}
\label{Fig:solution2}
\end{figure}

There remains to interpolate between the stick and slip parts of the dynamics.
To the best of our knowledge, this matching has not been done before for DRR-like equations, in relation with the idea of a finite time singularity of the slow dynamics.
According to matching theory, the solution in
the two interpolation domains (before and after the catastrophe) has to satisfy the set of equations common to the two phases, slow and fast,
  \begin{equation}
 \dot{u} = v
 \mathrm{,}
 \label{eq:DR1.resc.4}
 \end{equation}
 together with equations (\ref{eq:DR3.resc}) and (\ref{eq:DR2.resc.1}). The integral curves of (\ref{eq:DR3.resc}) and (\ref{eq:DR1.resc.4})  are solutions of the single linear ODE
 \begin{equation}
 \frac{{\mathrm{d}}\theta}{{\mathrm{d}}u} = \epsilon \theta + (1 +\epsilon) \xi u
\mathrm{,}
 \label{eq:racor1}
 \end{equation}
that gives
 \begin{equation}
 \theta(u) = c e^{\epsilon u} - \frac{(1+\epsilon)\xi}{\epsilon^2} (1 + \epsilon u)
\mathrm{,}
 \label{eq:raccord}
 \end{equation}
with   $c$ arbitrary
constant. In our problem this expression agrees well with the numerics, with $c$ calculated from a particular value of (${u,\theta}$) belonging to the matching region, see the discussion below concerning the figures (\ref{Fig:solution2}-a-b) and (\ref{Fig:solution3}).
The general time dependent problem in the matching region may be written as  $\dot{u} = \exp(\Phi(u))$ with
\begin{equation}
\Phi(u) = -c e^{\epsilon u}+ \frac{\xi}{\epsilon} u + \frac{(1+\epsilon)\xi}{\epsilon^2}
\mathrm{.}
 \label{eq:dvg}
 \end{equation}

The solution $ t - t_0 = \int_{u(t_0)}^{u} \mathrm{d}u'  \exp(-\Phi(u'))$ yields a singularity at finite time in the forward time direction if $\int_{u_0}^{\infty} \mathrm{d}u'  \exp(-\Phi(u'))$ is a converging integral, which requires $c$ to be negative. This singularity occurs at the end of the stick interval, as illustrated in figure (\ref{Fig:solution2}-a). However we must note that the solution (\ref{eq:dvg}) was derived within the adiabatic approximation (\ref{eq:stick}), which limits its validity to a domain we are going to precise. On the surface $S$, the time derivative of  $v=\exp(-\xi u-\theta)$ writes $\dot{v} = e^{-2 (\xi u + \theta)} \left[ \xi - \epsilon \theta - (1+\epsilon) u\right]$ by  using equation(\ref{eq:DR1.resc.4}) and the second equation (\ref{eq:2D}). Therefore in the space ($u,\theta$) the straight line of Cartesian equation
\begin{equation}
 \epsilon \theta(u)= (1+\epsilon) u-\xi
\mathrm{.}
 \label{eq:acc0}
 \end{equation}
splits the SM into a part such that trajectories
crossing this line
 tends to still lower values of $\dot{v}$, and so follow even more closely the SM, although when the crossing is in the other direction,
  $\dot{v}$ tends to increase, as well as $v$ itself (always positive) and so ends up with the finite time divergence just described.  A rough estimate of condition (\ref{eq:stick}) limits the 2D flow validity to the domain located above the line
 \begin{equation}
 \theta(u) = -\xi u-\ln(\gamma)
\mathrm{.}
 \label{eq:limitadiab}
 \end{equation}

\begin{figure}[htbp]
  \centerline{$\;\;$
 \includegraphics[height=1.80in]{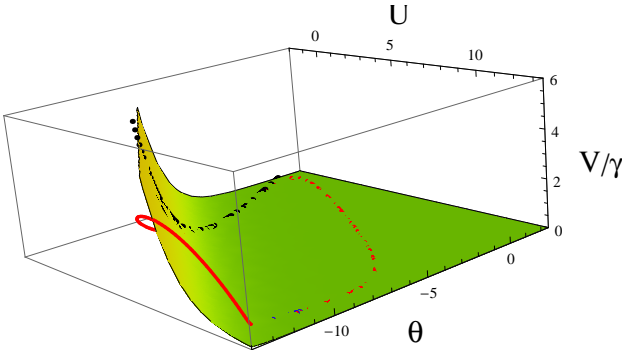}
 $\;\;$}
\caption{
 3D view of the trajectory (solid red curve) leaving and landing out and in the SM (yellow-green surface), the black part diverging on S displays the 2D flow in its "non slow" regime.
}
\label{Fig:solution4}
\end{figure}

These results are illustrated in figure (\ref{Fig:solution2}-b).
 In this figure the closed curve (blue-dashed) displays the limit cycle (3D flow) evolving counterclockwise in the phase space $(u,\theta)$. Superimposed onto this curve, we plot the stick episode (solid red portion), and the matching analytical  solutions (\ref{eq:raccord}) (orange and green curves). The solid straight line corresponds to the zero acceleration relation (\ref{eq:acc0}), joining the two extrema of the closed curve $\theta (u)$. The adiabatic approximation domain lies above the dashed straight line, equation (\ref{eq:limitadiab}). The two portions of analytical solutions fit well the two intermediate regimes, they are calculated with two distinct values of $c$ (derived from equation (\ref{eq:raccord}) and using values of $(u,\theta)$ before and after the catastrophe respectively).
The orange curve which belongs to the SM, matches the stick regime and the fast one, then it diverges on the surface $S$, while the limit cycle escapes from $S$. This happens when the adiabatic approximation fails.
The escape from the SM of the 3D flow in the fast regime is visible on figure  (\ref{Fig:solution4}).

 The physics of the oscillation can be understood as follows. The acceleration of the block  at the end of the stick phase is unstable because it lowers the friction and so feeds itself. The transition to the "sliding" fast regime occurs when the acceleration becomes big enough to make the inertia of the block relevant. In this fast regime the dynamical system gets one more dimension (in other terms the effect of inertia increases by one the number of coupled ODE's). If one neglects in this fast regime the friction, the dynamics is the one of a harmonic oscillator making a half swing while its velocity keeps  the same sign. When this oscillator has slowed down enough to yield back its energy to the potential energy of the spring, its velocity returns to small values and friction becomes so large that the SM is reached again. This could explain that, in earthquakes, a finite fraction of the initial elastic energy is not dissipated in the event itself, because part of it remains as potential energy after the large scale pendulum-like motion.
 In this model, the spring is under dilative strain before the slip episode and under compressive strain at the end of it.

\section{Lyapunov analysis along the trajectory}
\label{sec:Lyap}

We have noted above that the SM is everywhere attractive, taking argument of the negative sign in the r.h.s. of equation (\ref{eq:DR2.resc}). This statement  is invalid for the close vicinity of the SM, more precisely at distance of order $1/\gamma^2$ from the SM. In the very slow part of the trajectory, we may investigate how attractive is the SM by performing the linear stability analysis of the flow. Along a peculiar trajectory ($u_0(t),\ln(v_0(t)),\theta_0(t)$) the jacobian matrix is
 \begin{equation}
\left(
  \begin{array}{ccc}
    0 & 1 & 0 \\
    -\gamma^2 & -\frac{\gamma^2}{\xi v_0} & -\frac{\gamma^2}{\xi} \\
    0 & -A(\theta_0,v_0) & -v_0 \\
  \end{array}
\right)
 \mathrm{,}
  \label{eq:jacobian}
 \end{equation}
where

\begin{equation}
 A(\theta_0,v_0)= \theta_0+(1+\epsilon)(1+\ln(v_0))
\mathrm{.}
 \label{eq:A}
 \end{equation}

The eigenvalues of the matrix (\ref{eq:jacobian}) are solutions of the equation

\begin{equation}
 \lambda^3+(\frac{\gamma^2}{\xi v_0}+v_0)\lambda^2+\gamma^(\frac{1-A}{\xi}+1)\lambda+\gamma^2 v_0=0
\mathrm{,}
 \label{eq:lambda3}
 \end{equation}

\begin{figure}[htbp]
\centerline{$\;\;$
(a)\includegraphics[height=0.7in]{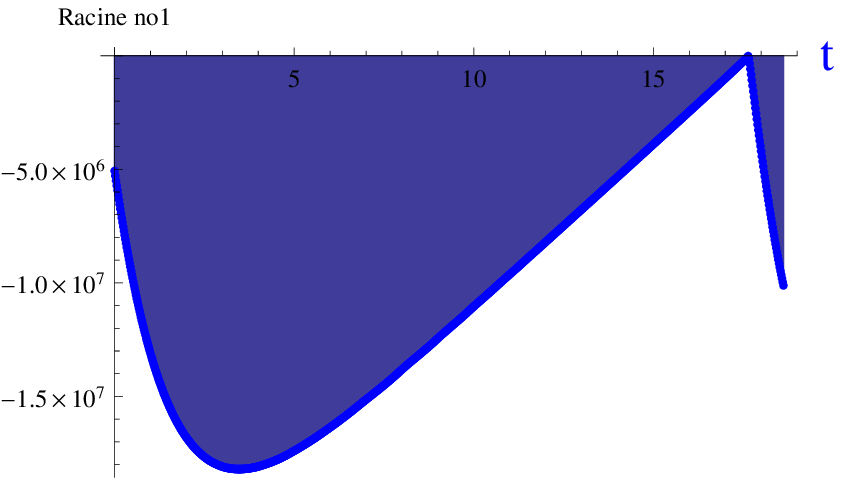}
(b)\includegraphics[height=1.2in]{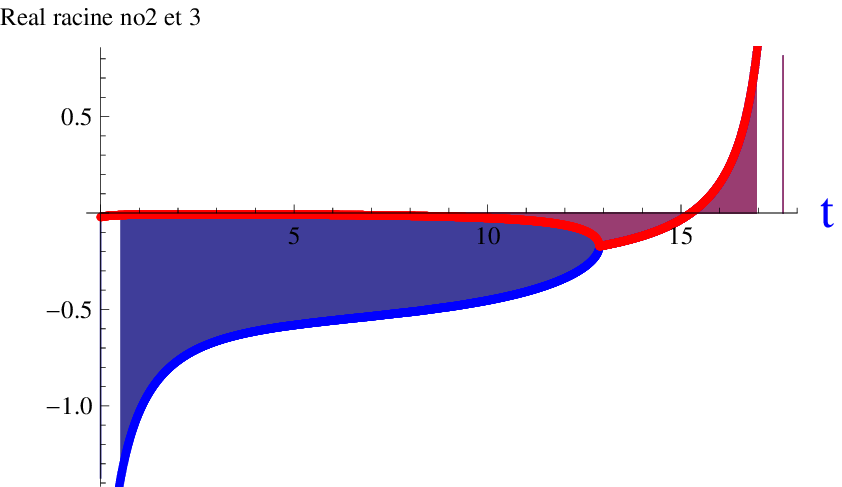}
 $\;\;$}
 \centerline{$\;\;$
(c)\includegraphics[height=1.2in]{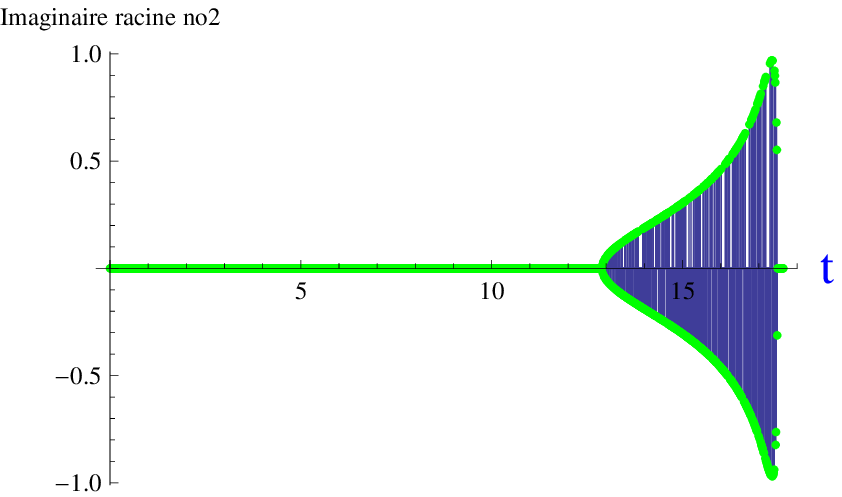}
 $\;\;$}
\caption{Lyapunov exponents along the trajectory, for the solution drawn in Figure (\ref{Fig:solution}-a). (a) negative real solution of equation (\ref{eq:lambda3}), (b)-(c) two other solutions, real parts in (b) and imaginary parts in (c). The real part of the two complex conjugate exponents crosses the zero real value at time $t_c -2.3t_0$, here $t_c=17.6475$.}
\label{Fig:Lyap}
\end{figure}

In the large $\gamma$ limit, the two eignevalues are

\begin{equation}
 \lambda(t)= -B(t)\pm \sqrt{B(t)^2-\xi v_0(t)^2}
\mathrm{,}
 \label{eq:lambda}
 \end{equation}
 with
 \begin{equation}
 B(t)= \frac{1}{2}v_0(1-A(\theta_0,v_0)+\xi)
\mathrm{.}
 \label{eq:lambda}
 \end{equation}
The three Lyapunov exponents are
drawn as  functions of time
 in figure (\ref{Fig:Lyap}), for the flow corresponding to figure (\ref{Fig:solution}-a), or (\ref{Fig:solution2}-b) or (\ref{Fig:solution4}). One of the eigenvalue (curve a) is negative all along the trajectory. The two other eigenvalues are real in the first part of the trajectory, and become complex conjugate at time about $t_c-4.6$, see figure (b). The real parts are the red and blue curves in (b) , the imaginary parts are drawn in figure (c).
   The real part crosses zero at time $t\sim t_c - 2.3$, with the eigenfrequency $\omega$ of order unity. This occurs
 inside the "very slow" stick regime  ending near $t \sim t_c-t_0$ where $v=1$. At this time, the Lyapunov analysis becomes invalid, because the motion cannot be considered as steady during the time interval $2\pi/\omega$.

\section{Response to noise}
\label{sec:correl}
\begin{figure}[htbp]
\centerline{$\;\;$
(a)\includegraphics[height=1.in]{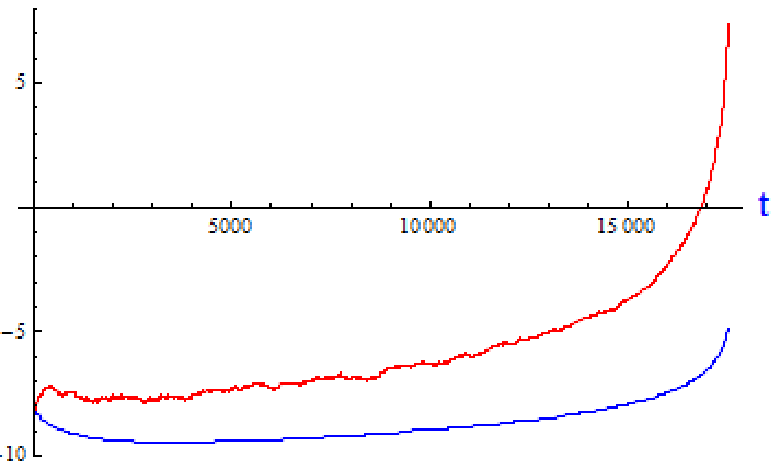}
(b)\includegraphics[height=1.2in]{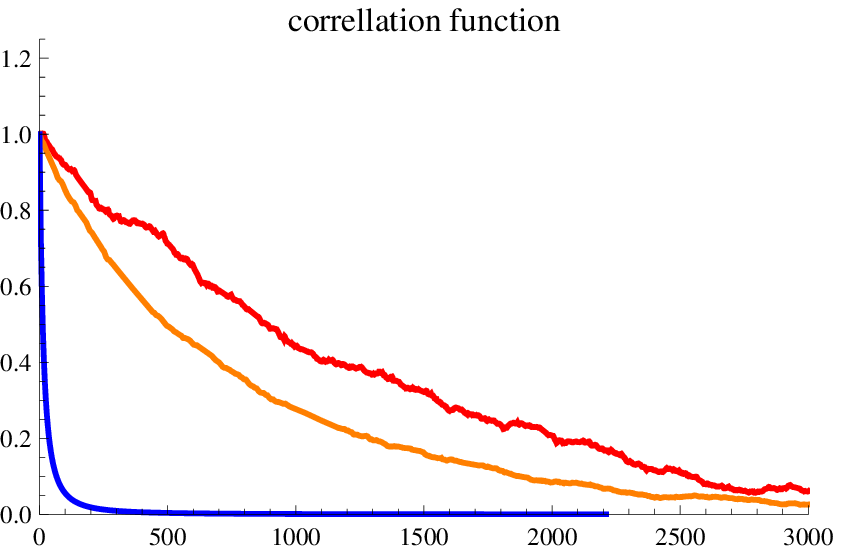}
$\;\;$}
\caption{Response to noise (a)Comparaison of the increase of velocity and its standard deviation, in log scale,versus $\gamma t$, until $tc-0.1$,
(b) correlation function $ \Gamma_u(t,t-t')$ as function of $t'$ for three time $t=t_c -2.3$ (red), $t=t_c-1$ (orange), $t=tc-30/\gamma$ (blue), for same parameters as in figure (\ref{Fig:solution}).
}
\label{Fig:sigcor}
\end{figure}

In view of forecasting catastrophes we have studied the response of the DRR system
 to an external white noise, as we did for the saddle-node bifurcation in the potential model \cite{SNL}.
 We solved the  set of stochastic equations

  \begin{equation}
 \left \{ \begin{array}{l}
 \dot{u} = v - 1 + \eta_u f_u(t)\\
 \dot{v} = - \gamma^2 \left(u + \frac{1}{\xi}(\theta + \ln(v))\right)+\eta_v f_v(t) \\
  \dot{\theta} = - v \left(\theta + (1+\epsilon) \ln(v)\right)+\eta_{\theta}f_{\theta}(t)
 \mathrm{,}
\end{array}
\right. \label{eq:withnoise}
\end{equation}
where  $f_u,f_v,f_{\theta}$ are three independent noise functions with short memory time (of order  $\delta t_{eqk}$) and $\eta_u,\eta_v,\eta_{\theta}$ are amplitudes smaller than $\gamma^2$. We have calculated the correlation functions of $x(t)$, $x$  standing for any function $u$, $v$, or $\theta$,

 \begin{equation}
 \Gamma_x(t,t-t')= <x(t)x(t-t')>-<x(t)><x(t-t')>
\mathrm{,}
 \label{eq:correl}
 \end{equation}

which are functions depending on the delay $t'$ but also on the time $t$, and the standard deviations
\begin{equation}
 \sigma_x(t)= \Gamma_x(t,t)
\mathrm{.}
 \label{eq:correl}
 \end{equation}
The results are very different from those of the potential model. First because here the standard deviation $\sigma_v$ of the response drastically increases with time before the event,  much more than the velocity $v(t)$ itself, that could be seen as a precursor. This contrasts with the potential case, where the standard deviation of the response grows nearly like $v(t)$ and therefore cannot be used as a precursor. The two functions $\sigma_v(t)$  and $v(t)$, are drawn in  Figure (\ref{Fig:sigcor}-a) in $\ln$ scale, showing that the growth of $\sigma_v$ is three order of magnitude larger than the growth of $v(t)$, on the time interval ($t_f-t_0, t_f$) , where the final time $t_f=t_c-0.1$ corresponds to about two months  before the event for typical earthquake case.

 Secondly the width of correlation function decreases very strongly during the large time interval $\sim  2t_0$ before the event, see (\ref{Fig:sigcor}-b). This could be understood as a sort of speed-up , contrary to what happens in the potential model  where the width  of $ \Gamma_u(t,t-t')$  becomes maximum shortly before $t_c$, in agreement with the well-known critical slowing-down phenomena (or critical opalescence in spatial systems).

\begin{figure}[htbp]
\centerline{$\;\;$
(a)\includegraphics[height=1.70in]{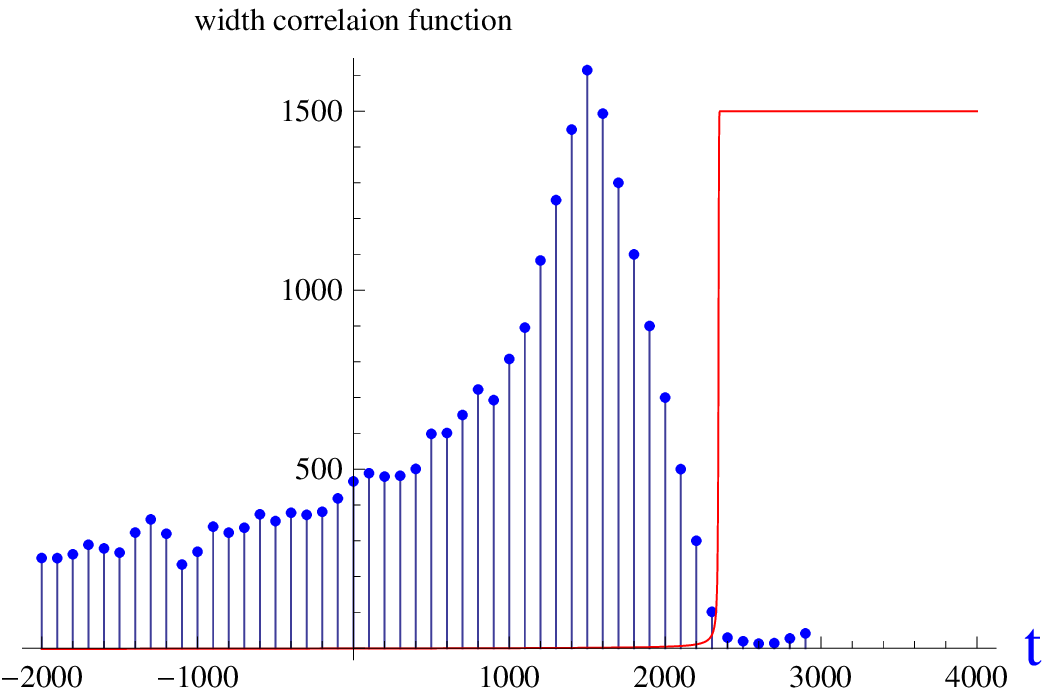}
 $\;\;$}
\centerline{$\;\;$
(b)\includegraphics[height=1.70in]{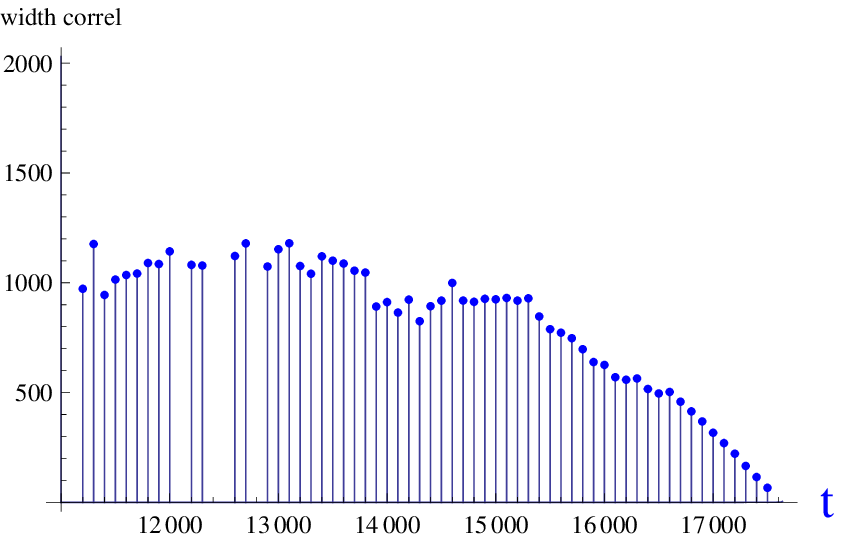} $\;\;$}
\caption{Correlation time for (a) the potential model in \cite{SNL},  (b) the present model with $\gamma=10^3$, $\xi=0.7$,$ \epsilon=0.72$}
\label{Fig:correl}
\end{figure}

This striking result is summarized in Figure (\ref{Fig:correl}) which displays the correlation time (mid-height width of correlation functions) of the fluctuating part of the displacement
in both cases, along the trajectory. For the saddle-node model (a-curve) the correlation time increases slowly
before it becomes maximum (critical slowing down) at time $ \sim t_c -t_0$ ($t_0$ being the intermediate time scale in this system\cite{SNL})
  then it drops abruptly
  before the catastrophe. For the DRR system we find that the correlation time of the response to noise (b-curve) decreases very slowly until the singularity (speed-up).

   The slow decrease also lasts a time interval of order $t_0$, but here
   \begin{equation}
 t_0/\delta t_{eqk}\sim \gamma
\mathrm{,}
 \label{eq:duree1}
 \end{equation}
    while it is
       \begin{equation}
 t_0/\delta t_{eqk}\sim \beta^{\frac{1}{3}}
\mathrm{,}
 \label{eq:duree2}
 \end{equation}
  for the saddle-node model. These two relations
     explain why the response to noise is changed much earlier in the DRR model than in the saddle-node model.
 Using parameters pertinent for earthquake, the precursor time $t_0$ is about few hours for the saddle-node model, while it is about few years for the DRR one.  Both results are related to the linear stability of the SM since the memory time of the response to external noise is maximum when the temporal integral of the largest stability exponent is close to zero, that gives a "precursor time" of order $t_0$ in both cases.

\section{Power law friction}
\label{sec:alfa}

Finally it is worth questioning the relevance of this idea of transition for other systems. In other terms how "generic" is this kind of transition? One can say first that it persists (as a way of going from slow to fast and conversely for the full dynamics) in the DRR equations by changing the numerical values of the parameters.  Another indication that it is a generic scenario is to check that it is still there in (slightly) modified equations.
  \begin{figure}[htbp]
\centerline{$\;\;$
(a)\includegraphics[height=1.in]{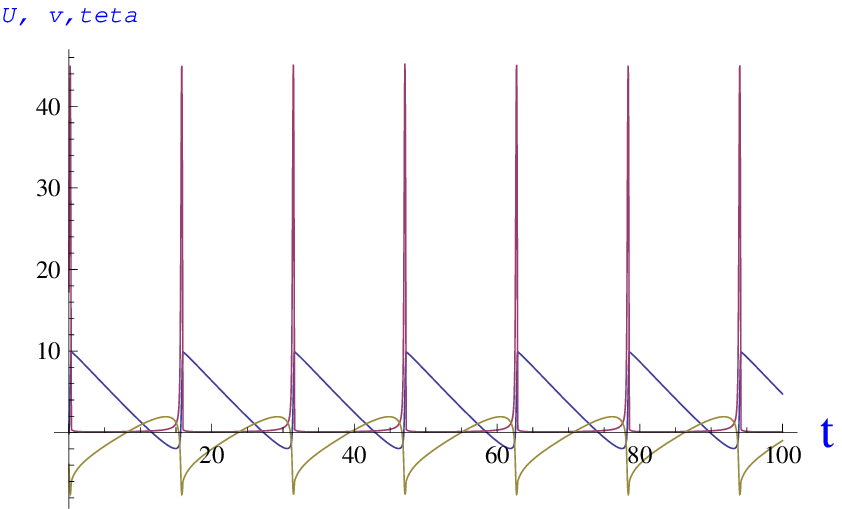}
(b)\includegraphics[height=1.in]{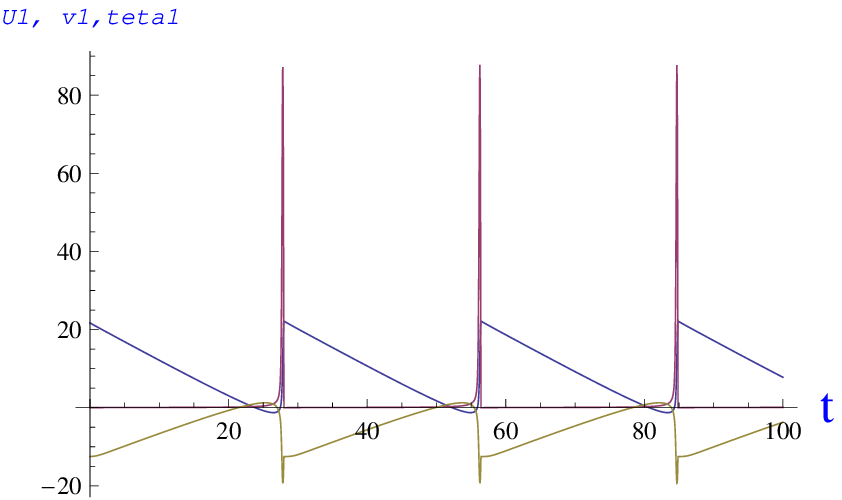} $\;\;$}
\centerline{$\;\;$
(c)\includegraphics[height=1.0in]{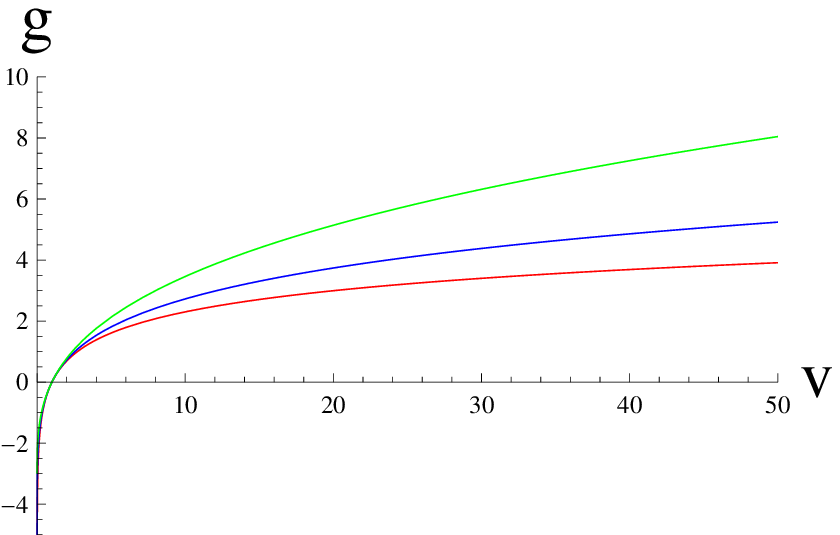}
(d)\includegraphics[height=1.0in]{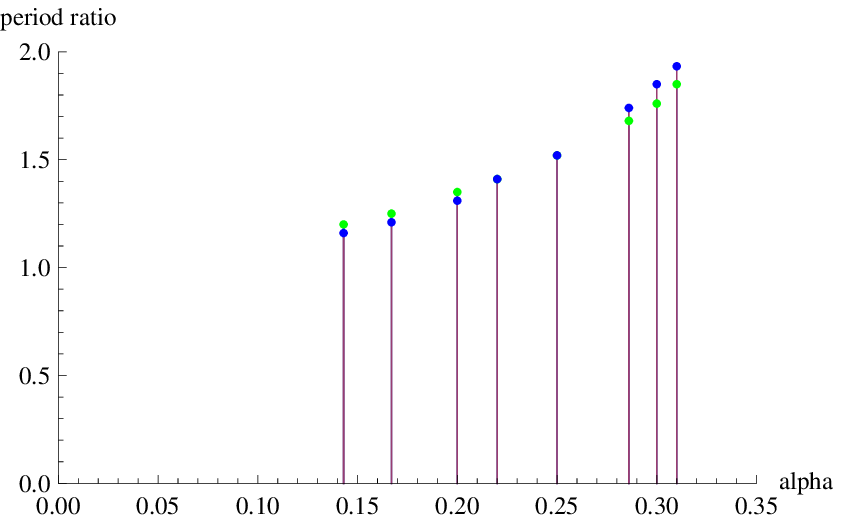}
 $\;\;$}
\caption{Solution of the DRR equations with $\ln(v)$ friction (a), and their modified form with a power law friction (b) ( equations (\ref{eq:puiss}) with $\alpha=0.31$) for the set of parameters $\gamma=10, \epsilon=1,\xi=0.7$. (c) $\ln(v)$ (red curve), and $ g_{\alpha}(v)$ (blue and green curves for $\alpha=0.143$ and $0.333$ respectively.
(d) The ratio of the limit cycle periods (blue points) and amplitudes (green points) for the power law and ln(v) cases, as function of $\alpha$}
\label{Fig:puiss}
\end{figure}

 We changed the logarithms dependence  $\ln(v)$ in the DRR equations (\ref{eq:DR2.resc}) and  (\ref{eq:DR3.resc}),  by the function
\begin{equation}
 g_{\alpha}(v) = \frac{v^{\alpha}-1}{\alpha} \\
\mathrm{,}
 \label{eq:eqg}
 \end{equation}
with $\alpha$ positive.
Solving the modified system of equations
 \begin{equation}
 \left \{ \begin{array}{l}
 \dot{u} = v - 1\\
 \dot{v} = - \gamma^2 \left(u + \frac{1}{\xi}(\theta + g_{\alpha}(v)\right)\\
  \dot{\theta} = - v \left(\theta + (1+\epsilon) g_{\alpha}(v)\right)
  \mathrm{,}
\end{array}
\right. \label{eq:puiss}
\end{equation}
 we found stable periodic solutions, with slow and fast steps. More precisely this appears above a certain threshold (close to the Hopf bifurcation the limit cycle has ordinary behavior). The curves (a) and (b) of figure (\ref{Fig:puiss}) display slow-fast limit cycles  obtained for the same parameter values, with the the $\ln(v)$ friction and the power law friction, respectively (DRR equations and equations (\ref{eq:puiss}) respectively). In both cases we observe that the period of the limit cycle increases with $\frac{\epsilon}{\xi}$ (for a given value of $\gamma$). Moreover in the power law case we find that the period of the limit cycle increases with $\alpha$, see figure (c).
 This occurs of because the nonlinearity increases with the parameter $\alpha$.
  The intermediate time (interval between the instant where $v(t)=1$ and $t_c$) is found to be nearly equal to unity,
 \begin{equation}
 t_0 \simeq1
  \mathrm{,}
 \label{eq:to}
\end{equation}

 The  jacobian matrix is
  \begin{equation}
\left(
  \begin{array}{ccc}
    0 & 1 & 0 \\
    -\gamma^2 & -\frac{\gamma^2}{\xi g'_0} & -\frac{\gamma^2}{\xi} \\
    0 & -A(\alpha) & -v_0 \\
  \end{array}
\right)
 \mathrm{,}
  \label{eq:jacobian2}
 \end{equation}
where $f'_0= v_0^{\alpha-1}$, $A(\alpha)=\theta_0 +(1+\epsilon)\frac{v_0^\alpha(\alpha+1)-1}{\alpha}$, and $\theta_0(t), v_0(t)$ are the dynamical variables along the trajectory, supposed to evolve very slowly with respect to the inverse of the imaginary part of any eigenvalue of the matrix(the Lyapunov  exponents). We found that the eigenvalues of matrix (\ref{eq:jacobian2}) and (\ref{eq:jacobian}) display similar behavior with one negative real exponent, and two others alternately real and complex. For example for $\gamma=10^3$,  $\xi=0.7$ and $\epsilon=0.78$  we observe that the real part of the complex conjugate exponents cross zero  at a time $t_c-2.45$, which is very similar to the result obtained with the $\ln(v)$ friction term, cf figure (\ref{Fig:Lyap}).

\begin{figure}[htbp]
\centerline{$\;\;$
\includegraphics[height=1.2in]{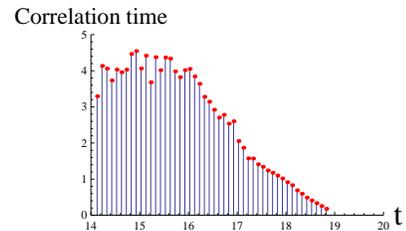}
 $\;\;$}
\caption{Half-height width of the correlation function, as a function of time for $\alpha=0.15, \gamma=10^3,\epsilon=0.78,\xi=0.7$}
\label{Fig:correl2}
\end{figure}

We have also studied the response to noise in the case of a power-law friction (equations (\ref{eq:puiss}) plus noise). The figure (\ref{Fig:correl2}) displays a progressive speed-up, the correlation time decreasing during an interval of order few $t_0$ before $t_c$, as in the case of the $\ln(v)$ friction law (compare with Figure (\ref{Fig:correl}-b)).

\section{Summary}
\label{sec:resume}

We have shown on a model of stick-slip dynamics that RO can result from a breakdown of the adiabatic approximation. In such systems
 the transition from slow to fast motion is just the result of an unchecked acceleration of the slow motion, a rather natural effect that should appear beyond solid/solid phenomena.  In the full DRR model, this acceleration is stopped by inertia, neglected in the adiabatic limit. In other contexts, the role of inertia could be played by feedback effects not taken into account in the equations of slow dynamics. Concerning the possibility of forecasting the fast event, and following ideas presented in \cite{SNL}, one finds that the response to an external source of noise is completely different in this model of stick-slip dynamics from what it is in the standard RO models. In the latter case the response to noise increases in amplitude before the transition and drifts to low frequencies (slowing-down). In stick-slip dynamics, the acceleration along the SM is also accompanied by a growth of the amplitude of the response, but it drifts to large frequencies (speeding-up).

\subsection{acknowledgment}

Paul Clavin is greatly acknowledged for discussions and interest for this work


\begin{thebibliography}{let1}
\bibitem{saddle} "Saddle folding" is a way to describe the slow to fast transition in dynamical system like the van der Pol equation in the strongly non linear regime. Near the saddle-fold the trajectory leaves the slow manifold at a point where the velocity field (which is directed along the variable having the large time derivative) becomes tangent to it. There the slow manifold  shows a "fold" (with respect to this large velocity direction). In the
case of equation (1) it occurs at $x=\pm 1$, in
 between (for $-1 < x <+1$) the trajectory becomes horizontal in the phase space ($x,y$) and the SM "folds".
\end{thebibliography}
	  \end{document}